\documentclass{ws-p8-50x6-00}

\usepackage{bm}

\begin{document}

\title{
Final state interactions in $\bm{\omega}$ photoproduction near threshold}

\author{Yongseok Oh}

\address{Institute of Physics and Applied Physics,
Yonsei University, Seoul 120-749, Korea
\\E-mail: yoh@phya.yonsei.ac.kr}

\author{T.-S. H. Lee}

\address{Physics Division, Argonne National Laboratory, Argonne,
Illinois 60439, U.S.A. \\
E-mail: lee@theory.phy.anl.gov}


\maketitle

\abstracts{
Vector meson photoproduction and electroproduction have been suggested
as a tool to find or confirm the nucleon resonances.
In order to extract more reliable informations on the nucleon resonances,
understanding the non-resonant background is indispensable.
We consider final state interactions in $\omega$ photoproduction as a
background production mechanism.
For the intermediate states, we consider nucleon--vector-meson
and nucleon-pion channels. The role of the final state interactions is
discussed in $\omega$ meson photoproduction near threshold.
}

\section{Introduction}

Vector meson production off nucleons near threshold attracts recent
interests in connection with the so-called ``missing resonance
problem'' \cite{CR00}.
By studying various physical quantities of vector meson production
one hopes to have information on the nucleon resonances especially
which couple rather strongly to the vector-meson--nucleon channel.
There have been recent progress to obtain such informations by
studying the processes of vector meson photoproduction
\cite{ZLB98c,OTL01,Zhao01,sasha}.
Among light vector mesons, $\omega$ photoproduction is studied in more
detail due to its simple isospin character \cite{OTL01,Zhao01}.

In order to extract information on the nucleon resonances from vector
meson production, it is essential to first understand the background
production mechanisms \cite{OTL00}.
As the background non-resonant production amplitudes one considers the
Pomeron exchange, one-boson exchange, and the nucleon pole terms.
Then the gap between the theoretical predictions on the background
production and the experimental data, e.g. in total and differential
cross sections, are expected to be explained by the terms including
nucleon resonances.
After adjusting the resonance parameters, other physical quantities,
especially polarization asymmetries, are predicted to have more
conclusive evidence for the nucleon resonances and it has been shown
that some polarization asymmetries are really sensitive to the presence
of nucleon resonances because of different helicity structure of the
production amplitudes.

However, in order to have more conclusive clues on the nucleon
resonances, it is essential to understand the background non-resonant
amplitudes in more detail.
We can have lessons from the study on the non-resonant part of
pion photoproduction and pion-nucleon scattering, which shows that the 
final state interactions are important to improve meson
exchange models \cite{NBL90,SL96}.
Such dynamical studies are important not only in understanding the
structure of the nucleon resonances but also for unitarity of the
scattering amplitude.
Therefore it is legitimate to improve the existing models by imposing
unitarity condition.

Such investigations are, however, intricate and many informations are
still unavailable to do a reliable model study.
In this work, therefore, we study final state interactions in $\omega$
photoproduction as our first step to construct a dynamical model for
vector meson photoproduction near threshold.
Because of its complexity, we only consider some one-loop diagrams in
this work that seem to be non-negligible in the production amplitude.
In the next Section, we discuss the non-resonant amplitude at tree level
and our method to compute the final state interactions for several
selected intermediate channels.
The preliminary numerical results are given in Sec. III with
discussions.

\begin{figure}[t]
\centering
\epsfig{file=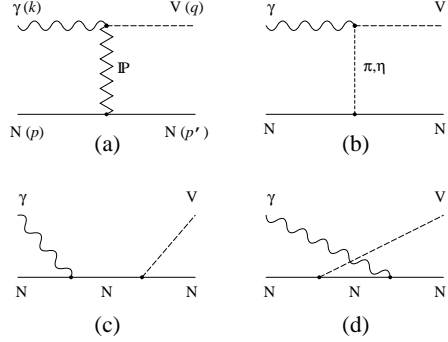, width=6cm}
\caption{Non-resonant interactions for $\omega$ photoproduction at tree
level. Here $V$ stands for the $\omega$ meson.}
\label{fig:tree}
\end{figure}

\section{Model}

As the non-resonant production process for $\omega$ photoproduction at
tree level, it is widely used to include the Pomeron exchange, $\pi$ and
$\eta$ exchanges, and the nucleon pole terms as depicted in
Fig.~\ref{fig:tree} \cite{ZLB98c,OTL01,Zhao01,sasha}.

However, it is well-known that the amplitudes obtained at tree level such
as in Fig.~\ref{fig:tree} do not satisfy unitarity.
For example, the soft Pomeron model of Donnachie and Landshoff
\cite{DL84} that has been used in analyzing $\omega$ photoproduction
has intercept larger than $1.0$ and hence violates the Froissart-Martin
bound \cite{Froi61,Mart63} that is a consequence of unitarity and the
partial wave expansion.
Imposing the unitarity condition to the process has been emphasized in
many respects \cite{MY66,SS68,MPP01}.
Unitarity condition is also crucial in developing dynamical models for
pion photoproduction and pion-nucleon interactions \cite{NBL90,SL96}.
Therefore it would be necessary to construct a unitarized model for
vector meson photoproduction near threshold to search for the
``missing nucleon resonances'' by, for example, solving the coupled-channel
equations, which would require very complicated calculations.
Before tackling to the unitarization of the amplitude directly, we first
compute final state interactions by considering several selected
intermediate channels.
It is the purpose of this work to compute several one loop diagrams in
$\omega$ photoproduction.

\begin{figure}[t]
\centering
\epsfig{file=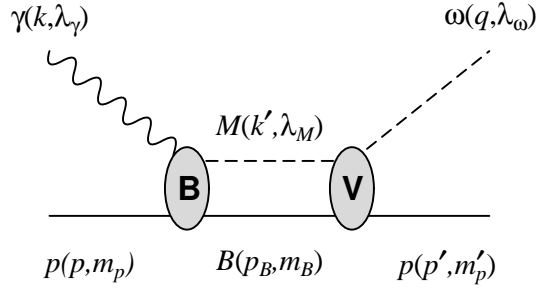, width=8cm}
\caption{Diagram for final state interactions for $\omega$
photoproduction.}
\label{fig:loop}
\end{figure}

Following Refs.~\cite{NBL90,SL96}, what we consider is the diagram shown
in Fig.~\ref{fig:loop}, which defines the momenta of the interacting
particles and their helicities/spins.
Then the amplitude shown in Fig.~\ref{fig:loop} is written as
\begin{equation}
(\mathrm{FSI})_{BM} = \int d {\bf k}' \frac{
\langle {\bf q}; \lambda_\omega m_p' | \mathsf{V} |
{\bf k}'; \lambda_M^{} m_p'' \rangle
\langle {\bf k}'; \lambda_M^{} m_p'' | \mathsf{B} |
{\bf k} \lambda_\gamma m_p \rangle}
{W - E_B({\bf k}') - E_M({\bf k}') + i \varepsilon},
\label{eq:fsi}
\end{equation}
after 3-dimension reduction, where $E_{B,M}({\bf k}) = \sqrt{ M_{B,M}^2
+ {\bf k}^2}$.
The intermediate baryon and meson masses are denoted by $M_B^{}$ and
$M_M^{}$, respectively, and $\lambda$ ($m$) is the helicity
(spin) of the particle.
Three-dimensional reduction of the full amplitude is not unique
\cite{HYL01} and we follow Refs.~\cite{NBL90,SL96} to obtain Eq.
(\ref{eq:fsi}).
It can be further decomposed into the principal integration part and the
delta function part as
\begin{eqnarray}
(\mathrm{FSI}) &=& 
\mathcal{P} \int dk' k'{}^2 \frac{\mathsf{V} (q,k';W)
\mathsf{B}(k',k)}{W - E_B(k') - E_M(k')}
\nonumber \\ && \mbox{}
-i \rho_{BM}^{} (k_t) \mathsf{V}(q,k_t;W) \mathsf{B}(k_t,k)
\theta(W - M_B^{} - M_M^{}),
\end{eqnarray}
where
\begin{equation}
\rho_{BM}^{}(k) = \frac{\pi k E_B(k) E_M(k)}{[ E_B(k) + E_M(k) ]},
\end{equation}
and $\theta(x)$ is the step function ($\theta(x) = 1$ for $x > 0$ and
$0$ otherwise).
The delta function part, which contains $\theta(x)$, arises when the
intermediate particles are their on mass shell and hence the
intermediate state on-shell momentum $k_t$ is defined by
\begin{equation}
W = E_B^{}(k_t) + E_M^{}(k_t).
\end{equation}

\begin{figure}[t]
\centering
\epsfig{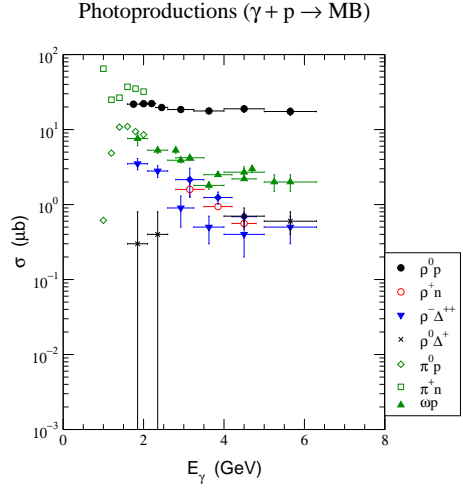}
\caption{Collected data for the total cross sections for meson
photoproduction. The data are from Ref. \protect\cite{Durham}.
The cross sections for $\pi^0 p$ and $\pi^+ n$ photoproductions are
from Ref. \protect\cite{DGL01} based on the SAID program.}
\label{fig:photon}
\end{figure}

The intermediate states contain various baryon-meson states allowed by
symmetries and quantum numbers.
Therefore, we start first by selecting the intermediate states that are
expected to be rather important.
For this purpose, we collect some experimental informations on the cross
sections of various meson photoproduction and the available experimental
data are shown in Fig.~\ref{fig:photon}.
One can see that the cross sections of $\pi$ and $\rho$ photoproductions
are considerably larger than the other reactions.
Based on this observation, we consider the intermediate $\pi^+ n$,
$\pi^0 p$, and $\rho^0 p$ states.

\begin{figure}[t]
\centering
\epsfig{file=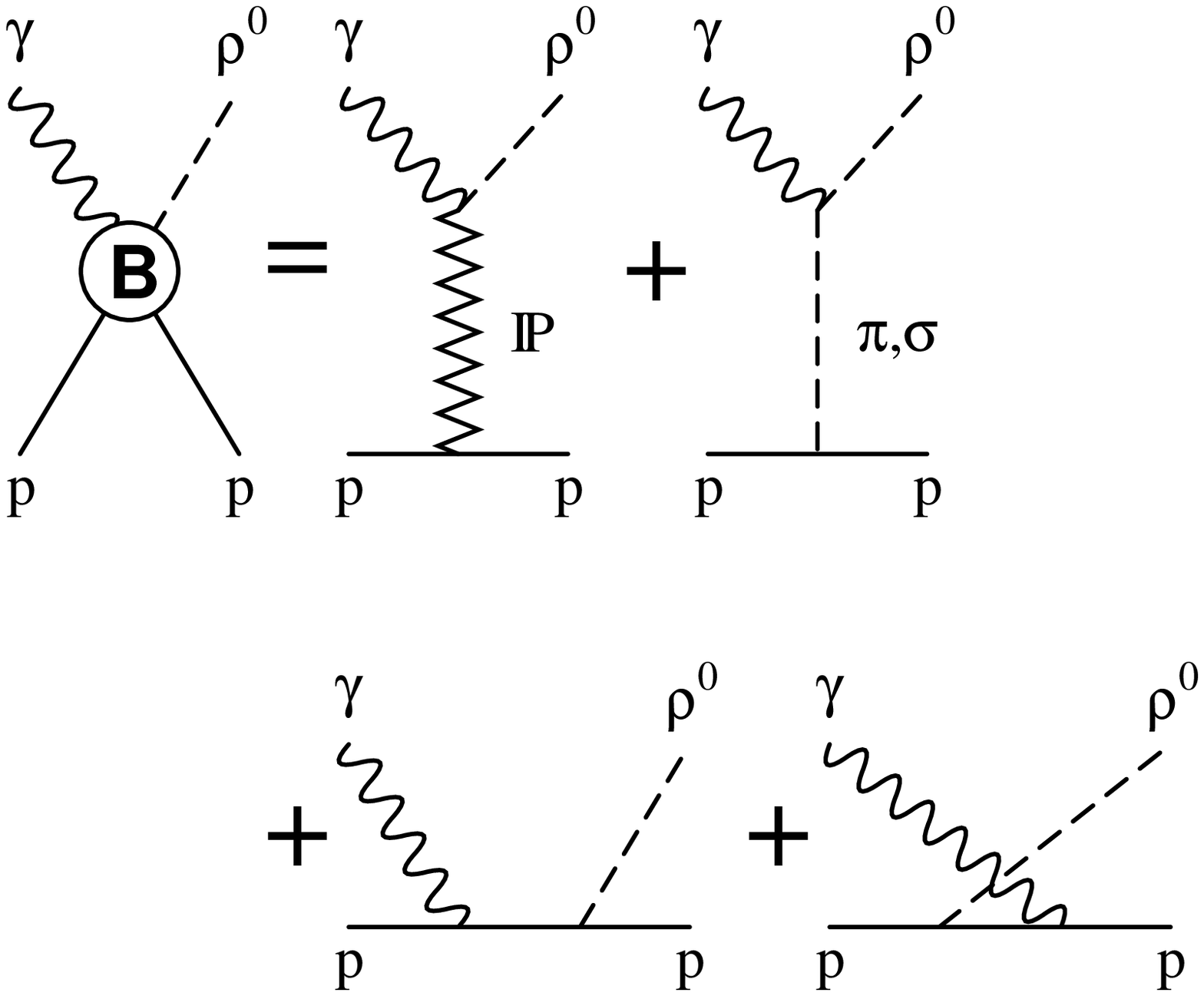, width=5.5cm} \qquad
\epsfig{file=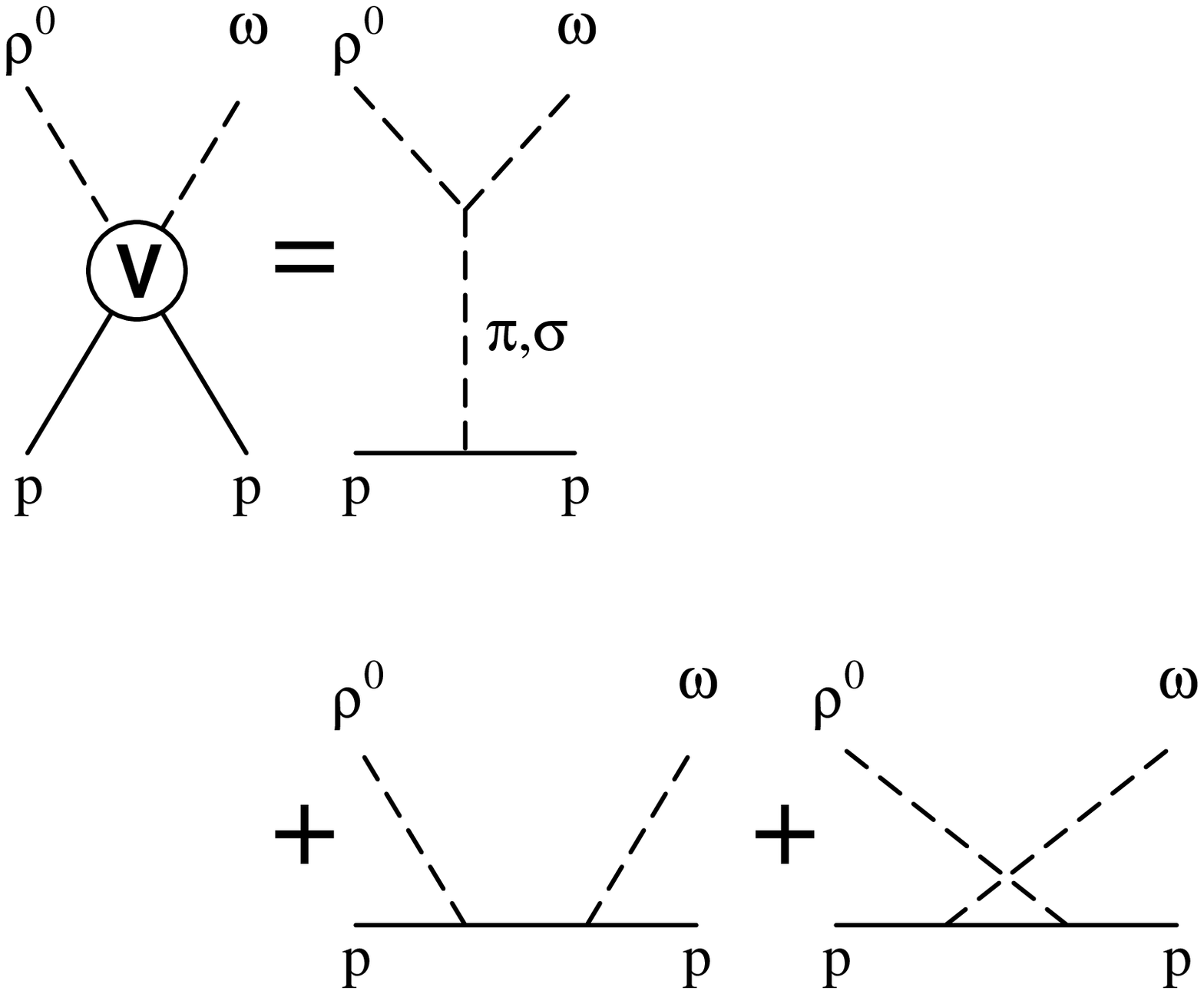, width=5.5cm}
\caption{$\omega$ photoproduction with intermediate $\rho^0$-$p$ state.}
\label{fig:rho-p}
\end{figure}

We first consider the intermediate $\rho^0 p$ state, which is depicted
in Fig.~\ref{fig:rho-p}.
Through the studies on $\rho$ photoproduction, we have learned that the
$\sigma$ meson exchange is important at low energies \cite{OTL00,FS96}.
Thus our amplitude for $\gamma p \to \rho^0 p$ contains the Pomeron
exchange, $\pi$ exchange, $\sigma$ exchange and nucleon pole terms as
given in Ref.~\cite{OTL00}.
The amplitude for $\rho^0 p \to \omega p$ is closely related to the
$\omega$ photoproduction amplitude at tree level that is given in Fig.
\ref{fig:tree} via vector meson dominance.
In addition to vector meson dominance, what we need is the
$\omega\rho\pi$ interaction Lagrangian, which reads
\begin{equation}
\mathcal{L}_{\omega\rho\pi}
= \frac{g_{\omega\rho\pi}}{2} \varepsilon^{\mu\nu\alpha\beta}
\partial_\mu \omega_\nu \mbox{ Tr } (\partial_\alpha \rho_\beta \pi),
\label{Lomega}
\end{equation}
where $\varepsilon^{0123} = +1$, $\pi = \bm{\pi} \cdot \bm{\tau}$, $\rho
= \bm{\rho} \cdot \bm{\tau}$.
The coupling constant $g_{\omega\rho\pi}$ was estimated by vector meson
dominance, massive Yang-Mills approach, and hidden gauge symmetry
approach, etc \cite{KKW96,KRS84,JJMP88,KVR01,FKTU85} and the estimates
are within $10 \sim 15$ GeV$^{-1}$.
In our study, we use
\begin{equation}
g_{\omega\rho\pi} = 12.9 \mbox{ GeV}^{-1},
\end{equation}
where its sign is fixed by SU(3) flavor symmetry.
Therefore our amplitude for $\rho^0 p \to \omega p$ contains the pion
exchange and the nucleon pole terms.

\begin{figure}[t]
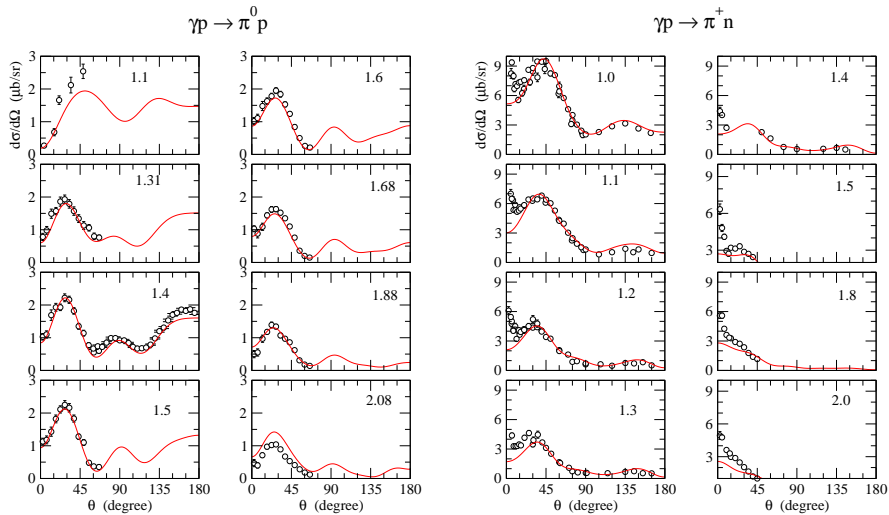

\centering
\epsfig{file=pi0.eps, width=5.5cm} \hfill
\epsfig{file=pip.eps, width=5.5cm} \hfill
\caption{Differential cross sections for $\gamma p \to \pi^0 p$ and
$\gamma p \to \pi^+ n$. The numbers in figures represent $E_\gamma$.
The experimental data are from Ref.~\protect\cite{Durham} and the solid
lines are from the calculation of Ref.~\protect\cite{DGL01}.}
\label{fig:pi}
\end{figure}

For intermediate pion-nucleon channel, we need to know the non-resonant
part of pion photoproduction.
But the most meson-exchange models are constructed to focus on the lower
energy region, so its direct extension to our energy region,
$W \simeq 2$ GeV, is quite questionable.
Because of this reason, we use the SAID program for the pion
photoproduction amplitudes.
Actually there is no experimental data for the total cross sections for
pion photoproduction and the data shown in Fig.~\ref{fig:photon} are
{\it not} experimental data but are extracted from the SAID program based
on Ref. \cite{DGL01}.
There can be a few comments on this method.
First, the SAID program is established to be valid up to $E_\gamma = 2$ GeV
and the extrapolation to the higher energy cannot be guaranteed.
Therefore we will use the program only in the limited energy region $W
\le 2$ GeV.
Second, the SAID program is believed to represent the experimental data.
This means that the extracted amplitude should be assumed to include all
nucleon resonance effects.
However, since there is no simple meson-exchange model for pion
photoproduction within our energy region, we will use this amplitude
keeping its limitation in mind.
The comparison of the model and the data are given in Fig.~\ref{fig:pi}.

\begin{figure}[t]
\centering
\epsfig{file=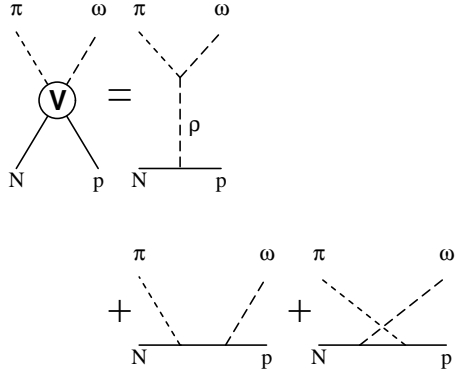, width=6cm}
\caption{A model for $\pi N \to \omega p$ reaction.}
\label{fig:pi-omega}
\end{figure}

\begin{figure}[t]
\centering
\epsfig{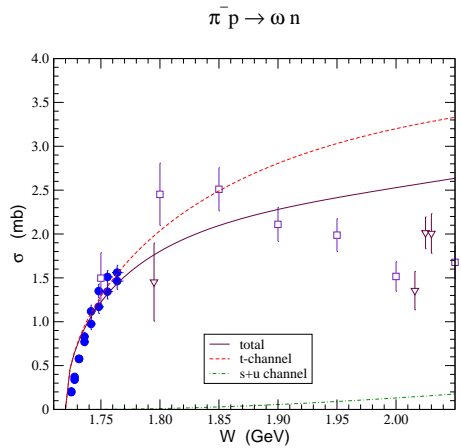}
\caption{Total cross section for $\pi^- p \to \omega n$.}
\label{fig:pip}
\end{figure}

We next need the amplitude for pion induced $\omega$ production.
This reaction has been discussed in Refs. \cite{PM01,LWF99,TKR01,PM01a}
recently.
It was also recently claimed that the final state interactions including
nucleon resonances are very crucial to explain the experimental data in
Ref. \cite{PM01a}.
Following  Refs. \cite{TKR01,PM01a}, in this study we use the
model shown in Fig.~\ref{fig:pi-omega} concentrating on the low energy
region, which is consistent with those model studies.
In addition to the $\omega\rho\pi$ interaction Lagrangian (\ref{Lomega}),
we need the $\omega NN$ and the $\rho NN$ couplings.
As in our previous studies \cite{OTL01}, we use
\begin{equation}
g_{\omega NN} = 10.35, \qquad \kappa_\omega = 0, \qquad g_{\rho NN} =
6.12, \qquad \kappa_\omega = 3.1,
\end{equation}
where the $\rho$-nucleon coupling is consistent with the $\rho$ coupling
universality.
Then the scattering amplitude includes the isospin factor, which reads
\begin{equation}
C_I^{} = \left\{ \begin{array}{l}
+1 \qquad \mbox{ for $\pi^0 p \to \omega p$} \\
\sqrt{2} \qquad \mbox{ for $\pi^- p \to \omega n$, $\pi^+ n \to \omega
p$} \\
-1 \qquad \mbox{ for $\pi^0 n \to \omega n$}
\end{array}
\right.
\end{equation}
The form factors of the vertices can be found, for example, in
Ref.~\cite{TKR01}.
The calculated total cross section for $\pi^- p \to \omega n$ is shown
in Fig.~\ref{fig:pip}.%
\footnote{
The experimental data for the cross section of this reaction near
threshold is controversial \cite{PM01a,HK99}.
In Fig.~\ref{fig:pip} we follow Ref. \cite{PM01a}.}

\section{Results and Discussions}

\begin{figure}[t]
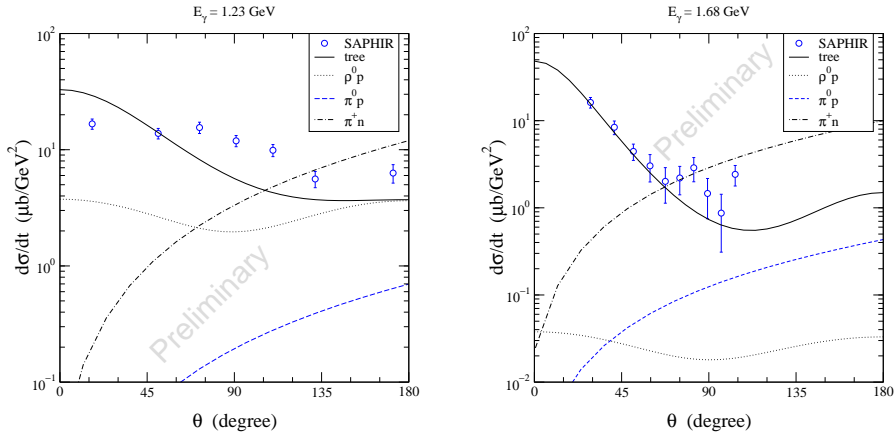

\centering
\epsfig{file=1.23.eps, width=5.5cm} \hfill
\epsfig{file=1.68.eps, width=5.5cm}
\caption{Differential cross sections for $\gamma p \to \omega p$ at
$E_\gamma = 1.23$ GeV (left panel) and $1.68$ GeV (right panel).
$\theta$ is the scattering angle in the center-of-mass frame.
The experimental data are from SAPHIR \protect\cite{Klein96}.}
\label{fig:fsi}
\end{figure}

With the amplitudes discussed so far, we first compute the differential
cross section for $\omega$ photoproduction at $E_\gamma = 1.23$ GeV and
$1.68$ GeV.
The preliminary results are shown in Fig.~\ref{fig:fsi}.
Since the purpose of this calculation is a rough estimate on the role of
the final state interactions, we do not try to adjust parameters to fit
the experimental data in this calculation.
In Fig.~\ref{fig:fsi}, the solid lines are the results at tree level
\cite{OTL01}, the dotted lines are from the intermediate $\rho^0 p$
state.
The intermediate $\pi^0 p$ and $\pi^+ n$ channels are represented by
dashed and dot-dashed lines, respectively.

As we expected from Fig.~\ref{fig:photon} the contributions from these
channels are not suppressed compared to the tree level results.
Especially the charged pion channel gives the most important
contribution among the intermediate state considered here.
The role of the neutral pion intermediate state seems to be smaller
than the other channels in the considered energy region.
Since the $\omega$ photoproduction cross section is at the same order
magnitude as the neutral pion photoproduction cross section and smaller
than the $\rho$ photoproduction cross section, the contribution from
the intermediate $\omega p$ channel is expected to be smaller than that
from the $\rho p$ channel.

We also note that the differential cross section from the
intermediate pion channel strongly depends on the momentum transfer $t$.
While the $\rho p$ channel differential cross sections do not strongly
depend on the scattering angle $\theta$, the pion channel contribution
gives rise to strong peaks at backward angles like the $u$-channel nucleon
exchange.
Thus, careful analyses are required to distinguish the two mechanisms at
backward scattering angles.

In summary, we calculate the one loop contribution to $\omega$
photoproduction with selected intermediate states.
We found that the final state interactions, especially charged pion
intermediate state, are not negligible in the considered energy region
and may affect the parameters of the nucleon resonances which will be
extracted from the forthcoming experimental data.
Since the polarization asymmetries are suggested to be the most useful
tools to investigate nucleon resonances, it would also be important to
check the contribution from the final state interactions to the
polarization asymmetries.

\section*{Acknowledgments}

Y.O. is grateful to Prof. M. Fujiwara for the warm hospitality during
the symposium.
This work was supported in part by the Brain Korea 21 project of Korean
Ministry of Education, the International Collaboration Program of
KOSEF under Grant No. 20006-111-01-2, and U.S. DOE Nuclear Physics
Division Contract No. W-31-109-ENG-38.

\end{document}